\documentstyle[11pt,aaspp4,tighten]{article}
\def\today{\rightline{\ifcase\month\or
        January\or February\or March\or April\or May\or June\or
        July\or August\or September\or October\or November\or December\fi
        \space\number\day, \number\year}}
\def\etal{{\it et al.}\ }

\def\s-1{{\rm\,s^{-1}}}

\def\spose#1{\hbox to 0pt{#1\hss}}
\def\C3H2{{\rm\,\rm C_3H_2}}
\def\NH3{{\rm\,\rm NH_3}}
\def\HOCO+{{\rm\,\rm HOCO^+}}
\def\lta{\mathrel{\spose{\lower 3pt\hbox{$\mathchar"218$}}
     \raise 2.0pt\hbox{$\mathchar"13C$}}}
\def\gta{\mathrel{\spose{\lower 3pt\hbox{$\mathchar"218$}}
     \raise 2.0pt\hbox{$\mathchar"13E$}}}

\begin{document}

\font\twelvei = cmmi10 scaled\magstep1 
       \font\teni = cmmi10 \font\seveni = cmmi7
\font\mbf = cmmib10 scaled\magstep1
       \font\mbfs = cmmib10 \font\mbfss = cmmib10 scaled 833
\font\msybf = cmbsy10 scaled\magstep1
       \font\msybfs = cmbsy10 \font\msybfss = cmbsy10 scaled 833
\textfont1 = \twelvei
       \scriptfont1 = \twelvei \scriptscriptfont1 = \teni
       \def\mit{\fam1 }
\textfont9 = \mbf
       \scriptfont9 = \mbfs \scriptscriptfont9 = \mbfss
       \def\bmit{\fam9 }
\textfont10 = \msybf
       \scriptfont10 = \msybfs \scriptscriptfont10 = \msybfss
       \def\bmsy{\fam10 }

\def\etal{{\it et al.~}}
\def\eg{{\it e.g.}}
\def\ie{{\it i.e.}}
\def\lsim{\raise0.3ex\hbox{$<$}\kern-0.75em{\lower0.65ex\hbox{$\sim$}}} 
\def\gsim{\raise0.3ex\hbox{$>$}\kern-0.75em{\lower0.65ex\hbox{$\sim$}}} 
\today
\title{A SURVEY FOR INFALL MOTIONS TOWARD STARLESS CORES.\\
III. $\rm CS~(3-2)$  and $\rm  DCO^+~(2-1)$ OBSERVATIONS}

\author{Chang Won Lee$^{1,2}$, Philip C. Myers$^{2}$, \& Ren\'e Plume$^{2,3}$}
\vskip 0.2in
\affil{$^1$Korea Astronomy Observatory,}
\affil {61-1 Hwaam-dong, Yusung-gu, Daejon 305-348, Korea}

\vskip 0.2in
\affil{$^2$Harvard-Smithsonian Center for Astrophysics,}
\affil {60 Garden Street, MS 42, Cambridge, MA  02138, USA}

\vskip 0.2in
\affil{$^3$University of Calgary, Department of Physics \& Astronomy}
\affil {2500 University Drive NW, Calgary, Alberta, Canada T2N 1N4}
 
\affil{E-mail: cwl@trao.re.kr, pmyers@cfa.harvard.edu, plume@ism.ucalgary.ca}

\vskip 1in
\begin{abstract}
We present  CS(3-2) and  DCO$^+$(2-1) observations of 94 starless cores
and compare the results with previous CS(2-1) and $\rm N_2H^+$(1-0) observations  
to study inward motions in starless cores.  
Eighty-four cores were detected in both CS and DCO$^+$ lines. 
A significant number of CS(3-2) profiles  and a small number of $\rm DCO^+$(2-1) lines
show the classical  ``infall asymmetry'' similar to that seen in CS(2-1) observations.  
The $\rm DCO^+$(2-1) lines, however, usually show a single Gaussian peak. 
The integrated intensity of $\rm N_2H^+$ correlates well with that of $\rm DCO^+$(2-1),
but poorly with that of CS(2-1) and CS(3-2), suggesting that CS suffers significantly 
more depletion onto grains than do either  $\rm DCO^+$ or $\rm N_2H^+$.
Despite these depletion effects, there is evidently enough optical
depth for the CS(3-2) and CS(2-1) spectral lines to exhibit infall asymmetries.

The velocity shifts of the CS(3-2) and (2-1) lines with respect to $\rm N_2H^+$
correlate well with each other and have similar distributions.
This implies that, in many cores, systematic inward motions of gaseous material may occur 
over a range of density of at least a factor $\sim 4$.

We identify 18 infall candidates based on observations of CS(3-2), CS(2-1), 
$\rm DCO^+$ (2-1) and $\rm N_2H^+$ (1-0).  The eight 
best candidates, L1355, L1498, L1521F, L1544, L158, L492, L694-2, and L1155C-1, 
each show at least four indications of infall asymmetry and no counter-indications.

Fits of the spectra to a 2-layer radiative transfer model in ten infall candidates 
suggest that the median effective  
line-of-sight speed of the inward-moving gas is $\sim 0.07$ $\rm km~s^{-1}$
for CS (3-2) and $\rm \sim 0.04~km~s^{-1}$ for CS(2-1).  
Considering that the optical depth obtained from the fits is usually smaller in CS(3-2)
than in (2-1) line, this may imply that CS(3-2) usually traces inner denser gas in higher
inward motions than CS(2-1). 
However, it is also possible that this conclusion is not representative of 
all starless core infall candidates, due to the statistically small number analyzed here.
Further line observations will be useful to test this conclusion. 
\end{abstract}

\keywords{ISM: Globules; ISM: Kinematics and Dynamics; Stars: Formation }

%\begin{slocitlum}{2}
\clearpage

\section{Introduction}

Starless cores are dense ($\rm \ga 10^4~cm^{-3}$) condensations in molecular clouds 
without any embedded Young Stellar Objects (YSOs) (e.g., Lee \& Myers 1999-LM99).  
They are the best known candidates to possibly form stars 
(Tafalla et al. 1998; Lee, Myers, \& Tafalla 1999-LMT99; 
Jijina, Myers, \& Adams 1999; Lee, Myers, \& Tafalla 2001-LMT01) and,
 therefore, provide an opportunity to study  the initial conditions
of star formation (Benson \& Myers 1989; Ward-Thompson et al. 1994; 
Shirley et al. 2000; Andr\'e, Ward-Thompson, \& Barsony 2000).  

The inward motions in starless cores are one of the essential elements needed to understand 
the onset of star formation.
These motions can be  studied by observing two kinds of molecular lines,
optically thin and thick tracers, to detect the spectral ``{\it infall asymmetry}'' :
a double peaked profile in which the blue component is brighter than the red one, or 
a skewed single blue peak in an optically thick tracer (e.g., CS2-1 and $\rm H_2CO$2-1), 
and a symmetric single peak in an optically thin tracer 
(e.g., $\rm C^{18}O$1-0 and $\rm N_2H^+$1-0) (Hummer \& Rybicki 1968; Leung \& Brown 1977). 

This method has been used  quite successfully to observe some general features of 
inward motions in starless cores (e.g., LMT99, LMT01).  
The inward motions in starless cores  are found to be 
slow (typically 0.05  - 0.1 $\rm km~s^{^-1}$), and spatially extended (typically 0.1 pc) (LMT01) 
and can be partly explained by ambipolar diffusion in a super-critical core, or by
turbulent dissipation (e.g., Nakano 1998; Myers \& Lazarian 1998; Ciolek \& Basu 2000). 
However, the physics of inward motions in starless cores
is still not fully understood. 

Different molecular species and various transitions of specific molecules are 
sensitive to different chemical and physical conditions  and, therefore,
can probe different physical regions of the infalling cores.
Therefore, observations 
of many different species and transitions would help to widen our present
knowledge of inward motions in starless cores
(e.g., Onishi et al. 1999; Caselli et al. 1999;  Gregersen \& Evans 2000; Lee et al. 2003).

This study is a continuation of our previous systematic infall studies, 
which used CS(2-1) and $\rm N_2H^+$(1-0) by LMT99 and LMT01.  
In this study we use two $\it new$ tracers: CS(3-2) and $\rm DCO^+$(2-1).
The CS(3-2) line has higher critical density ($\rm n_{cr}\approx 1.3\times10^{6}~cm^{-3}$)
than CS(2-1) by  a factor of about 4 (Evans 1999) and so is likely a much better tracer of 
dense gas. 
Thus, the CS(3-2) line will be a better probe of the kinematics closer 
to the nucleus of the core unless the optical depth of the CS(3-2) is larger 
than CS(2-1).  DCO$^+$(2-1)  is  similar to $\rm N_2H^+$(1-0);
it is also a  high density tracer and is usually optically thin.
However, the DCO$^+$(2-1) has high enough opacity and is bright enough
to make extensive surveys feasible.
In this paper we present new survey results with CS(3-2) and DCO$^+$(2-1) toward starless cores and,
by combining these with previous CS(2-1) and $\rm N_2H^+$(1-0) data, discuss 
how inward motions in starless cores occur.

In $\S$2 we explain details of the observations such as the observational equipment 
and methods, target selection, frequency determination of observing lines,
and data reduction. 
Detection statistics, spectral features of CS and DCO$^+$, $\rm \delta V$ analysis 
of the observing lines, and selection of infall candidates  are described in $\S$3.
Implications of the observational results are discussed in $\S$4. 
In $\S$5 we summarize the main conclusions of this study. 

\section{Observations}

	We observed both the CS(3-2) and the DCO$^+$ (2-1) transitions at
the NRAO 12-m telescope\footnote[1] 
{The National Radio Astronomy Observatory is a facility of the National Science Foundation, 
operated under cooperative agreement by Associated Universities, Inc.} in two seasons, 
1998 June and October.
Typical system temperatures were  200 - 300 K in good weather. 
The dual channel 2 mm receivers were used to simultaneously observe two polarizations 
to obtain a final spectrum with better signal-to-noise (S/N) 
by averaging the two polarization states.
Two different autocorrelators were used during the two seasons.
The hybrid autocorrelator operating with a 12.5 MHz bandwidth was used for the observations 
in June and was replaced by the millimeter autocorrelator (MAC) with a 200 MHz bandwidth
for the observations in October. Both correlators  give spectra with a
24 KHz resolution, corresponding to $\rm \sim 0.05~km~s^{-1}$ at 145 GHz. 
The beam size (FWHM) of the 12-m telescope is $43''$ at 145 GHz. 
The spectra are presented in $\rm T_R^*$ - antenna temperature corrected for 
atmospheric attenuation, radiative loss, and rear-ward and forward scattering and spillover. 
The corrected main beam efficiency $\rm \eta_{mb}^*$ (0.85 at 147 GHz) is used 
to convert from the $\rm T_R^*$ scale to the main beam brightness temperature scale. 
Integration time varied from about 10 minutes to about 1 hour among sources and resulted in
S/N ratios greater than 20 for most sources.   
Frequency switching mode was used with a 3.1 MHz throw for sky subtraction of spectra. 

Most targets were selected from LM99, particularly if they had
strong CS (2-1)  lines.  However, an additional fifteen sources with  
low declinations
(which were not observable from the Haystack observatory) are observed  
here for the first time.   The target list  is given in Table 1 with details of  
the source selection  noted in the last column.
The starless cores in this list are all within a few hundred pc, dense
(between $\rm 10^4 \sim 10^5~cm^{-3} $; LMT01), compact (optical sizes  
of $0.05\sim 0.35$ pc; LM99),
and have narrow line widths ($\Delta V_{\rm FWHM}$ of $\rm N_2H^+$ $\sim 0.2 -  
0.4~ \rm km~s^{-1}$; LMT99).

	Accurate frequencies for the molecular lines are always important 
in a quantitative study of inward motions in starless cores, 
since the lines  are usually very narrow (LMT99 \& LMT01).  
To determine our line frequencies we used molecular lines observed in L1512.
First, the $\rm C^{18}O~1-0$ line observed at the  IRAM 30-m (Tafalla et al. 2004 in prep.)
was used as a reference ($\nu_{rest} = 109782.173\pm 0.002$ MHz; LMT01). 
This was compared  with the $\rm DCO^+~2-1$ line profile obtained with the same telescope
(Caselli et al. 2004 in prep.). For quantitative comparison we used
the Gaussian fit results of two spectra to determine 
the amount of the frequency shift for the $\rm DCO^+~2-1$ profile to coincide 
with $\rm C^{18}O~1-0$ profile.
From this experiment we obtain a $\rm DCO^+~2-1$ line frequency of 
$\nu_{rest} = 144077.303\pm 0.002$ MHz.
The uncertainty is an error propagated in the process of deriving the value of 
the frequency shift to the determined frequency of $\rm DCO^+~2-1$ line.

The frequency of CS 3-2 was then determined by comparing our observed $\rm DCO^+$2-1 profile 
with the CS 3-2 profile.  
Since their spectral shapes are very close to  Gaussian, it was straightforward
to determine the frequency for the CS 3-2 line;  $\nu_{rest} = 146969.029\pm 0.002$ MHz.
Note that this frequency is in good agreement with the value independently determined
from laboratory measurements (Gottlieb, Myers, \& Thaddeus 2003- $146969.026\pm 0.001$ MHz).

All data reduction of spectral lines was performed 
using the `CLASS' reduction software (Buisson et al. 1994).

\section{Results}
\subsection{Detection Statistics}

Detailed results of our observations for all 94 targets 
are summarized in Table 1.
Ninety-one out of 94 sources were detected in CS(3--2).
Eighty-four sources were detected in both CS(3--2) and $\rm DCO^+~$(2-1) lines. 
All spectra are displayed in Fig. 1.
 
\subsection{CS Spectra}

In this section we describe the features of the CS(3-2) spectra and compare them with 
CS(2-1) line profiles from LMT99. 
As was true with CS(2-1), CS(3-2) was easily detected in nearly all the sources,
indicating that CS is a ubiquitous molecule in  cold molecular cores.
Most of the CS(3-2) profiles also show similar features to those seen in CS(2-1) and, similar 
to the CS(2-1) lines, the CS(3-2) lines  have distinct spectral line
shapes which can be classified as follows;

1. Two peaks, with the blue peak brighter than the red; e.g., L1521F, L1544, L1582A,
L1689B, L234E-S, L429-1, L429-1, L1155C-1, and L1063.

2. A blue peak with a red shoulder; e.g., L1355, L1498, L1399-1, L1445, 
TMC2, L1552, L183, L694-2, and L1148.

3. A single symmetric peak; e.g., B217-2, L1507A-1, L1517A, and L1512-1.

4. A red peak with a blue shoulder; e.g., L1399-2, and CB246-2. 

5. Two peaks with the red component brighter than the blue one.

6. Two peaks with similar brightness.

However, there are also some differences between the CS(2-1) and CS(3-2) line profiles.  
Notably,  the CS(2-1) line is often (for about 10 sources) seen to have broad wings 
which are not seen in CS(3-2). 
Spectra for L429-1, L183, and L1251 are such typical examples (Fig. 2). 
Another difference is that 
the CS(3-2) observations have no members in either group 5 
or group 6, while the (2-1) profiles have  
at least one core in group 5 and five cores in  group 6.

We also find some sources for which the classification, 
as determine by the CS(3-2) line, is  different from that determined by CS(2-1).
L1445, L183 (named as L183B in LMT99), and L694-2 which were placed in group 1 by the CS(2-1) 
observations
are now members of group 2. L234E-S and L1155C-1 which were determined to be in group 2
by CS(2-1) observations are now members of group 1.
CB246-2 which was in group 5  by CS(2-1) is now in group 4.
L429-1 and L1544 which were in group 6 by CS(2-1) now show asymmetric double
peaks characteristic of infall and so are now classified as members of group 1.
Finally, L1063 which had the typical symmetric profiles characteristic of group 3 in CS(2-1)
now shows infall asymmetry (group 1) in CS(3-2).

\subsection{DCO$^+$ Spectra}

$\rm DCO^+$(2-1) lines usually show a single Gaussian peak,
indicating that they are likely optically thin. 
However, there are also several sources with asymmetric $\rm DCO^+$ profiles caused by
higher optical depth. 
Such sources  are (1) L1521F, L1544, L1696B, L1709B-2, L158, L63, L492, and L694-2 
which show  spectra with either skewed infall asymmetry or a blue peak and red shoulder,
(2) L234E-S which shows a spectrum with a red peak and blue shoulder, 
and (3) L183 which shows a spectrum with two peaks, the red peak brighter than the blue peak.

The spectral features of DCO$^+$(2-1) and CS(3-2) vary quite a bit from source to source.
Therefore, we list sources in the following groups that highlight different 
combinations of CS(3-2) and $\rm DCO^+$(2-1) spectral features:
(a) sources with symmetric profiles in both lines, e.g., L1521B-2, B217-2, CB22, L1507A-1, 
CB23, L1517A, L1512-1, and CB130-1, (b) sources with asymmetric CS and
symmetric $\rm DCO^+$ profiles, e.g., TMC2, L1689B, L429-2, L1148, L1155C-1, 
and CB246-2, (c) sources with similarly asymmetric profiles in both lines, e.g., L1709B-2, 
L492, and L694-2, (d) sources with double peaked infall asymmetry in CS  and
single peaked infall asymmetry in $\rm DCO^+$, e.g., L1521F and L1544, and 
(e) sources showing opposite asymmetries in CS and $\rm DCO^+$,
e.g., L183, and L234E-S.    

There is a strong tendency for sources that show $\rm DCO^+$ infall asymmetry 
to also show CS  infall asymmetry. Of the eight such DCO$^+$ cores listed in 
group (1) above, five also show infall asymmetry in CS(3-2), 
while two (L1696B and L1709B-2) show no significant asymmetry in CS line.

\subsection{$\rm \delta V$ analysis for CS(3-2) and $\rm DCO^+$(2-1)}

The degree of spectral asymmetry in the CS profiles can be  quantitively measured 
by the normalized velocity difference [$\rm  \delta V_{CS32} = 
(V_{CS32}-V_{N_2H^+})/\Delta V_{N_2H^+}$] between CS(3-2) and 
N$_2$H$^+$(1-0) (Mardones et al. 1997).  
$\rm V_{CS32}$ is the velocity determined from a Gaussian fit to the
brighter spectral component of the CS(3-2) profile, after masking the less bright component or 
the skewed one by ignoring such parts of the profile.  
$\rm V_{N_2H^+}$ and $\rm \Delta V_{N_2H^+}$ are the velocity
and FWHM of N$_2$H$^+$(1-0) taken from LMT99.
Here $\rm V_{N_2H^+}$ is slightly changed from LMT99 
since we use the more accurate frequency of N$_2$H$^+$ as determined in LMT01.
Table 1 lists the line intensity and $\rm V_{CS32}$ for CS(3-2) in column 4 \& 6, and 
the derived $\rm \delta V_{CS32}$ in column 8.
Fig. 3 compares the distribution of $\rm  \delta V_{CS32}$ from our sample 
with that of $\rm  \delta V_{CS21}$ from LMT99 (slightly revised by using the 
frequencies determined in LMT01) and shows that $\rm  \delta V_{CS32}$ is significantly 
skewed to the blue side ($\rm \delta V_{CS} < 0$) and is similar in appearance 
to the $\rm  \delta V_{CS21}$ distribution.
In addition, the average ($\pm$ s.e.m -- standard error of the mean of the distribution) 
of $\rm \delta V_{CS32}$ is $-0.14\pm 0.03$  
which is also similar to the average of $\rm \delta V_{CS21}$ ($-0.13\pm 0.04$).
Finally Fig. 4 compares the $\delta V$'s 
of CS(3-2) with those of CS(2-1) for each source, showing a good correlation 
(correlation coefficient; $r=$0.79) between the two transitions.

We also derive the normalized velocity difference  ($\rm  \delta V_{DCO^+}$)
between $\rm DCO^+$(2-1) and $\rm N_2H^+$(1-0) (where the $\rm N_2H^+$ observations
are also taken from LMT99). 
Fig. 5 shows the distribution of $\rm  \delta V_{DCO^+}$ which appears to be quite 
symmetric, unlike the distribution seen in CS (Fig. 3).  
This implies that the statistical occurrence of 
inward motions in starless cores as traced by $\rm DCO^+$ is lower than it is for CS, which 
may be partially due to low optical depths in the $\rm DCO^+$ line.  

\subsection{Infall candidates}
The determination of infall candidates is one of most important objectives 
in this kind of survey.  With more molecular line data we can better identify infall candidates.
Therefore, in this section we combine the CS(3-2) and DCO$^+$ data from this survey with the 
 CS(2-1) and $\rm N_2H^+$ data of LMT99.
To identify infall candidates we use the following spectral properties (as used in LMT99);
(1) the values of $\delta V$ for CS(2-1), CS(3-2), and $\rm DCO^+$(2-1)
with respect to the $\rm N_2H^+$(1-0), (2) the intensity ratios ($\rm T_b/T_r$) of 
the blue to red components in double-peaked CS and $\rm DCO^+$ spectra, and 
the main component in $\rm N_2H^+$ spectra. These properties are flagged by ``color codes'':
1 for $\delta V \le -5\sigma_{\delta V}$ (a blue normalized velocity difference), 0 for 
$\vert \delta V\vert < 5\sigma_{\delta V}$ (neutral), and -1 for 
$\delta V \ge 5\sigma_{\delta V}$ (a red normalized velocity difference).
Property (2) is indicated by a 1 for $\rm T_b/T_r > 1+\sigma$ 
(blue peak stronger than the red peak), 
0 for $\rm 1-\sigma \le T_b/T_r \le 1+\sigma$ (peaks of equal strength), and 
-1 for $\rm T_b/T_r < 1-\sigma$ (red peak stronger than the blue peak), where 
the values of $\rm T_b$ and $\rm T_r$ were derived from
Gaussian fits to each component of the spectra with double peaks (or single
peak  with a distinct shoulder). 

Table 2 lists the results of the spectral properties with their associated numerical 
``color codes''.  These numerical values 
enable us to select infall candidates in a more quantitative fashion since, 
it is now possible to calculate the
total color code of each source simply as the algebraic sum of the individual color codes 
associated with each spectral property.
The last two columns in Table 2 give the total color code (C) and the total number (N) of 
spectra considered in the calculation of the total color code for each core.
The combination of this information allows us to more accurately identify infall candidates 
even between cores that may have the same value of C: a greater N indicates 
a slightly stronger infall candidate. 
To select infall candidates, we put a slight emphasis on the N
parameter. This is mainly because an increase in N usually corresponds to a detection of
N$_2$H$^+$, which is crucial to the $\rm \delta V$ analysis of
CS and DCO$^+$ and also allows us to set an additional flag for 
$\rm T_b/T_r$ in the analysis of the N$_2$H$^+$ main hyperfine line. 
In addition, detection of the
N$_2$H$^+$ line indicates the existence of dense gas
inside the source which increases the likelihood of star formation activity.
For example, although L1445 and L1689B are classified as
possible infall candidates with the same value of C (i.e. 2), we believe L1689B to be a 
slightly better infall candidate because it has more data (N=5) included in the color code
calculation (due to the detection of the high density tracer $\rm N_2H^+$).  
Therefore, with this rating scheme we can more accurately identify 
potential infall candidates than we could in LMT99.

Sources with C=2 or 3 have somewhat `blue' spectral properties, and
we define those as {\it possible infall candidates}.
Such candidates are {\bf L1445}, {\bf TMC2}, {\bf TMC1}, {\bf L1552},
{\bf L183}, {\bf L1696A}, {\bf L1689B}, {\bf L234E-S}, {\bf L429-1}, and {\bf L1234}.

Note that, while L1689B was classified as a strong infall candidate   
in LMT01, it is not in this paper.
The main reason is that this source is so weak in $\rm  N_2H^+$ that there is
a large uncertainty in the determination of $\rm \delta V$.  As can be  
seen in Table 1, L1689B has a large negative $\rm \delta V$,
but also has a large propagation uncertainty
(mainly a result of the low S/N of the $\rm  N_2H^+$ observations)
which gives $\rm \delta V$/$\sigma$ = -0.58/0.12 = -4.8.
If this source had been brighter, it almost certainly would have  
been classified as one of the strong infall candidates.
Again, this shows that the detection of N$_2$H$^+$ is an
important factor in the determination of infall candidates.
 
L234E-S is the same source as L234E-1 which was selected as infall candidate from LMT99.
Here L183 and L234E-S have a mixture of `blue' and `red' spectral features,
although they have a total color code of 2, meaning that they have different spectral shapes
in the different tracers.
We note that two ``possible'' infall candidates TMC1 and L1234 have C=2 due to their line
center shifts but they are unusual in that their profiles do not have any significant 
line asymmetry.
More studies will be needed to better understand physical situation of these sources.

Sources with C $>$ 3 are defined as {\it strong infall candidates} 
in the sense that nearly all their spectral properties are given `blue' flags.
Such sources are {\bf L1355}, {\bf L1498}, {\bf L1521F}, {\bf L1544}, {\bf L158}, 
{\bf L492}, {\bf L694-2}, and {\bf L1155C-1}. 
Note that all the infall candidates  suggested here have been previously selected by LMT99 and LMT01.  
However, L1524-4, B18-3, CB23, and L1622A-2 which were previously chosen 
as the infall candidates in LMT99
have been  dropped from the current list because they do not exhibit 
`blue' spectral features either in their value of $\rm \delta V_{CS32}$, 
or in the intensity ratio ($\rm T_b/T_r$) of the CS(3-2) profiles.  
Although L1521F is listed as a strong infall candidate from our rating  
system, it is a very complicated source and difficult to understand.
L1512F  shows typical infall asymmetry in high angular resolution  
CS(2-1) observations
(LMT99),  $\rm HCO^+$(3-2) \& (4-3) (Onishi et al. 1999), CS(3-2) (this  study), and
in the main component of $\rm N_2H^+$(1-0) (LMT01).  
However it also  shows extended red asymmetry in the CS(2-1) maps (LMT01) and 
a mixture of  `blue' and `red'
spectral features in the CS(3-2) maps (Lee et al. 2003 in prep.).
Further detailed observations using various tracers are being made to  
clarify  its kinematics (e.g., Shinnaga et al. 2004, Caselli et al. 2004 in prep.).

\section{Discussion}
\subsection{Infall speeds}
We derive infall speeds for a sub-sample of cores via fits to their CS spectra using 
a two-layer model consisting of  a cool (2.7 K) absorbing front screen 
moving away from us and a warm emitting rear layer approaching us 
with the same speed (see LMT01).  
Our non-linear least squares fitting routine has 5 free parameters, peak optical depth, 
infall speed, velocity dispersion, excitation temperature of the rear layer and LSR velocity.
In the model, both layers are assumed to have the same velocity dispersion and move
with the same speed but in opposite direction.
Note that the number of fit parameters is more than the 3 required to
fit any line with a Gaussian.  However, the two additional parameters are
needed to specify the peak optical depth and the infall speed. Thus,
this model has the least number of parameters possible to describe
a line with infall asymmetry due to self-absorption and internal motion.
Indeed, each parameter reflects different aspects
of the line profile and so can be well constrained by the fitting  
procedure.
The line width, LSR velocity, and the excitation temperature
are very well determined by the width, centroid velocity, and intensity  
scale of the observed spectrum.
The  peak optical depth is sensitive to the depth of
the self-absorption dip and so fits to spectra with a clear dip or red  
shoulder can easily constrain the optical depth.  
In addition, since the first four parameters can be constrained by fits to the  
spectra, the last parameter
- infall speed - can be reliably determined with good precision.
This is the main reason that, to derive infall speed,  we fit only spectra with clear  
double peaks,
or a blue peak and a clear red shoulder.  

Fits were conducted for the CS(3-2) profiles of 12 sources showing such infall asymmetry.
Fig. 6 shows that our model easily reproduces the observed spectral features.  
Note that in the figure we did not fit a model profile  to the CS(2-1) spectra of two sources 
(L1355 and L1155C-1) although they do show infall asymmetry. 
This is because they show neither clear 
double peaks nor a red shoulder in their spectra and so the fits to their spectra
would result in large errors in the derived infall speeds. 
Typical 1-D infall speeds for the sources modeled
are $\sim$ $\rm 0.02 - 0.13~km~s^{-1}$.

We also obtained the infall speeds for a sub-sample of 8 sources using
 fits to the CS(2-1) spectra which show similar asymmetry to that of CS(3-2).  
The resulting infall speeds derived from our model fits to both 
the CS(3-2) and CS(2-1) observations are compared in Fig. 7.
An interesting feature in Fig. 7 is that the CS(3-2) lines generally
have higher infall speeds than CS(2-1); typically $\sim 0.07$ $\rm km~s^{-1}$
for CS (3-2) and $\rm \sim 0.04~km~s^{-1}$ for CS (2-1).  

We believe that this is not dominated by fitting errors to the infall 
speeds since the difference between the CS(2-1) and CS(3-2) speeds
is much larger than the fitting error.
As discussed in LMT01, we tried to estimate the fitting error of 
the infall speed by taking two best-fit model spectra of CS(2-1) and CS(3-2)
for L1521F, adding random noise with the same rms as the observed noise, 
and re-fitting a model to the spectra. Fig. 8-(a) plots the distribution of infall speeds 
obtained from the best-fit for twenty synthetic CS(2-1) and (3-2) spectra in L1521F.
This figure shows 
that the two distributions of infall speeds are clearly separated. 
The estimated 1 sigma uncertainties of the distribution of infall speeds are
about $\sim 0.001 \rm ~km~s^{-1}$ for the CS(2-1) and $\sim 0.003 \rm ~km~s^{-1}$ 
for the CS(3-2), which are negligibly small compared to the difference 
($\sim 0.03 \rm ~km~s^{-1}$) between the infall speeds for two lines.
Note that most of spectra used in the fitting procedure are as good as the L1521F spectra.
Thus, their fit results are thought to be also as good as for L1521F, suggesting 
that the tendency for CS(3-2) line to trace higher inward speeds than CS(2-1)
is real.

One might question how the infall speed can be determined
with precision as high as indicated here. We note that the infall
speed, like the centroid velocity, is sensitive to emission from the
entire spectrum, so with sufficient signal to noise ratio one can
determine the infall speed to a fraction of a channel width. 
Note that this precision, due to random uncertainties, is not the 
same as the accuracy, due to systematic uncertainties. Detailed 
comparisons show that the accuracy of these estimates of two-layer 
infall speed, with respect to the values from a more sophisticated 
Monte Carlo model, is likely to be a few 0.01 $\rm km~s^{-1}$, 
for spectra with resolution and signal to noise
ratio similar to those considered here (De Vries and Myers 2004, in preparation). 
Furthermore the infall 
speed from the two-layer model usually increases monotonically with 
the infall speed from the detailed models. Thus 
even if the two-layer infall speed is systematically
incorrect by a few 0.01 $\rm km~s^{-1}$, the sign of its gradient with optical
depth, which we derive in this paper, is probably unaffected.

Figure 8-(b) and (c) are another test to demonstrate how different infall speeds 
are required to  produce the  difference
between the CS(2-1) and CS(3-2) line profiles.  The solid line in  
Figure 8-(b) shows the best  $\chi^2$ fit to the
CS(2-1) spectrum in L1521F, which produces an infall speed of $0.014  
\rm ~km~s^{-1}$ and a good match to the observed spectral line profile.  
Similarly, the solid line in  Figure 8-(c) shows the best  $\chi^2$ fit to the
CS(3-2) spectrum in L1521F, producing an infall speed of $0.045 \rm  
~km~s^{-1}$ and a  good fit
to the observed spectral line profile.   However, the dashed line in  
Figure 8-(b) shows the spectrum obtained if
we attempt to fit the CS(2-1) spectrum with the CS(3-2) velocity (and  
vice-versa in Figure 8-(c)).  Casual inspection
of these fits, as well as the $\chi^2$ values, reveals that small  
changes in the infall velocity result in poor fits to the spectra.
This strongly suggests that the
difference between the CS(2-1) and CS(3-2) infall velocities is  
statistically significant, and that the CS(3-2) line traces higher  
infall speeds than does the CS(2-1) line.

This analysis with the 2-layer model indicates that the CS 3-2 profiles 
require slightly but distinctly higher speeds than the 2-1 profiles.
However it still remains to be determined whether these higher 
speeds correspond to denser gas in the inner regions of the cores. 
Higher speeds associated with denser gas 
if the line formation were dominated by excitation effects, since the 
centrally  condensed structure of the core implies that the 
excitation temperature decreases outward, and since the 3-2 line has 
a higher critical density than the 2-1 line by a factor $\sim 4$ (Evans 
1999).  But the higher speeds could also correspond to less dense 
gas, if the line formation were influenced more by optical depth 
effects, and if the optical depth of the 3-2 line exceeds significantly 
that of the 2-1 line (e.g. Lucas 1976). 
The other possible problem in our analysis is resolution difference 
between CS(2-1) ($\sim 25''$) and (3-2) data ($\sim 43''$) which may affect
the profile shapes that we observe. 
To deal with these problems, we use the FCRAO data for CS(2-1) (LMT01) 
which have  similar beam resolution ($\sim 52''$) to that of CS (3-2) 
data. Using the same two layers model, we derive the infall speed and optical depths
for FCRAO CS(2-1) lines.
Table 3 lists the parameters obtained from the fits
for all the data, including 12m data (CS 3-2), Haystack and FCRAO data (CS 2-1). 
The table clearly indicates that, regardless of the different beam resolution of the data,
the CS(3-2) line tends to trace higher 
infall speed than the CS(2-1) line, and that the CS(3-2) lines have 
lower optical depth than CS(2-1) lines for all sources except for 
L158 and L1445.

We also believe that, as the case for the infall speeds, the optical depth 
derived from the fits is not dominated by fitting errors.
In performing the  error analysis for the infall speed in L1521F (see above), 
the optical depths for each of the twenty CS(2-1) and (3-2) spectra were obtained 
at the same time. Fig. 9-(a) shows the distribution of such optical depths,
indicating  that the two distributions of the optical depths are clearly separated. 
The estimated 1 sigma uncertainties of the distribution are
about $\sim 0.035$ for the CS(2-1) and $\sim 0.03$ 
for the CS(3-2), which are very small compared to the difference 
($\sim 0.24$) between the optical depths.
Fig. 9-(b) and (c) also confirm that the difference 
is  statistically significant as it was for the infall speeds (Fig. 8-b and c).

Fig. 10 visualizes the distribution of the sources in infall speed
difference versus optical depth difference between two CS(3-2) and (2-1) lines.
The data consist of seven points 
for FCRAO Data and eight points for Haystack data. 
Data points from the same core are joined by a line. 
As shown, nearly all of the data points (11 of 15) and majority of the cores 
(6 1/2 out of 10) lie in the upper right quadrant, 
the regime where infall speed increases inward,  
indicating that the 3-2 line  usually traces gas which is closer to 
the core center and which has faster infall speed than does the 2-1 
line.  This means  that the infall speed generally increases inward 
for this sample and these lines.  Note that this trend is evident in 
both the FCRAO and Haystack data, and so apparently the change in 
resolution between these telescopes does not reverse the tendency for 
points to occupy the upper right quadrant in Fig. 10.

In summary the conclusion that CS(3-2) traces faster and inner gas than CS(2-1) 
seems now reliable at least in infall candidates considered among our sample.
However, we are still cautious of whether this is necessarily true of most starless cores, 
or even of most infall candidates, because of the small number of objects involved 
in our analysis.
Further detailed observations of more infall candidates with the same lines are needed
to test whether the conclusion generally apply to starless cores or most infall candidates. 
Furthermore, our conclusion may depend on many factors such as observing lines 
and should be further tested.  For example infall 
speeds from these CS observations will be able to be compared with those from high-S/N 
observations of ``low-depletion'' species like  $\rm N_2H^+$ and DCO$^+$ (see the section
4.3 ) if they show enough infall asymmetry.  
Further such line observations are necessarily required.

\subsection{Implication of the similar $\rm \delta V$ distribution 
for CS(3-2) and CS(2-1)}
Fig. 3 and 4 provide important clues as to how inward motions 
occur.  
Given the expectation that CS(3-2) is 
a better probe of the denser inner region, one might expect 
a more skewed $\rm \delta V$ distribution for CS(3-2) than for CS(2-1).
However, the $\rm \delta V$ distribution for CS(3-2) is 
found to be statistically very similar to that for CS(2-1) (Fig. 3).  What does this
imply about inward motions in cores?
We saw in the previous section that CS(3-2) traces faster infalling
gas than CS(2-1) by typically $\rm \sim 0.03~km~s^{-1}$. 
The increase in infall velocity would result in a similar  
 increase in line velocity which was used to compute $\rm \delta V$.
Note, however, that this increase is not that great compared
to the $\rm N_2H^+$ line width, typically 0.3 $\rm km~s^{-1}$.  Therefore
the typical increase in $\rm \delta V$ is about 0.03/0.3, or 0.1, 
or only about half of a bin size in the
histograms in Fig. 3 which explains why the histograms look similar despite
the higher critical density and higher infall speed for  CS(3-2) compared to (2-1).  
The similar $\rm \delta V$ distribution 
between CS(3-2) and CS(2-1) may suggest that, in many cores,  inward-moving gas extends 
over a substantial range of core densities.
Furthermore, the good correlation between $\rm \delta V_{CS21}$ and 
$\rm \delta V_{CS32}$ (Fig. 4) suggests 
that the lower density and higher density gas probed by the 2-1 and 3-2 
lines respectively move together, in a systematic rather than random fashion.

\subsection{Depletion of CS and $\rm DCO^+$} 
The CS molecule is known to easily deplete out of the gas phase 
by adsorbing onto dust in  cold dense cores (e.g., Bergin \& Langer 1997).  
Therefore, it can be argued that CS cannot be used to probe 
the kinematics in the nuclei of infalling cores.   
$\rm N_2H^+$, however, does not seem to suffer from the effects of depletion 
as readily as CS (Tafalla et al. 2002; Bergin \& Langer 1997). 
Recent mapping surveys of several starless cores have shown that the $\rm DCO^+$ emission 
usually has a similar intensity distribution to that of $\rm N_2H^+$ 
(Lee et al. 2004 and Bourke et al. 2004 in prep.), suggesting that $\rm DCO^+$ also
does not readily deplete onto grains.
Fig. 11 supports this idea. 
The figure compares the integrated intensities of $\rm N_2H^+$,  
CS(2-1), and CS(3-2) of the sources.
Note that, in constructing  Fig. 11, we dropped all
sources with clear self-absorption features in CS(2-1) (17: number of the dropped 
sources), CS(3-2) (21) and DCO+(2-1) (14) to minimize any optical depth effects.
The middle and bottom panels of Fig. 11 compare the $\rm N_2H^+$ integrated
intensities with that of CS(3-2) and CS(2-1) and show weak to negligible correlations 
[$r=$0.54  between $\rm N_2H^+$ and CS(3-2),
and $r=$0.40 for $\rm N_2H^+$ and CS(2-1) ], suggesting the possibility 
that CS is depleting out onto grains.  
However, we should note that despite the strong likelihood of CS depletion 
in the cores we observe, there is evidently still enough optical depth,
excitation gradient, and relative motion to give the  CS (2-1) 
and (3-2) spectral lines an  infall asymmetry. 
It will need a more detailed chemical evolution models 
that self-consistently solve for radiative transfer effects
to specify the extent and degree of this depletion.

On the other hand, the upper panel of Fig. 11 compares the $\rm N_2H^+$  and 
$\rm DCO^+$ integrated intensities and shows a stronger correlation ($r=$0.75)
which suggests that the $\rm DCO^+$ does not suffer from significant depletion 
(or, at least, only depletes as much as $\rm N_2H^+$).  

\section{Summary}
We present the results of a systematic single pointing survey of CS(3-2) and  
DCO$^+$(2-1) emission in 
starless cores.  Eighty-four of the observed 94 targets were detected in both lines.
Most of the CS(3-2) profiles have similar spectral line shapes to those 
seen in CS(2-1) by LMT99 and are usually self-absorbed. 
Notably, the CS(2-1) is often seen to have the broader wings than the CS(3-2). 
On the other hand, the $\rm DCO^+$(2-1) lines usually show a single Gaussian peak,
indicating low optical depths.  There are, however, eight exceptional sources
showing infall profiles in $\rm DCO^+$(2-1). 
Using the spectral properties of CS(3-2) and DCO$^+$, together with CS(2-1) and $\rm
N_2H^+$(1-0) from LMT99, we introduced line asymmetry ``color codes"  to 
better identify infall candidates.  Eight strong infall candidates are:
L1355, L1498, L1521F, L1544, L158, L492, L694-2, and 
L1155C-1, and 10 possible infall candidates are:
L1445, TMC2, TMC1, L1552, L183, L1696A, L1689B, L234E-S, L429-1, and L1234.

We applied a two layer radiative transfer model to the CS spectra of the infall candidates 
to derive infall speeds and found  that CS(3-2) generally traces 
higher infall speeds than CS(2-1) by a small, but measurable amount 
(typically 0.03 $\rm km~s^{-1}$).  
The model fits generally give lower optical depth in CS(3-2) than in (2-1).
All together, the higher  CS(3-2) infall velocity may arise from denser gas 
closer to the core nucleus than (2-1).
However, because of small number of sample considered in this study,
we are rather cautious of whether this conclusion will apply for most of starless cores or
infall candidates. 
Further various line observations will be necessary to confirm or refute our conclusion.

We find evidence for CS depletion from a lack of correlation  
between the CS and $\rm N_2H^+$ (1-0) line intensities.  
However, the good correlation between $\rm DCO^+$ and $\rm N_2H^+$
intensities suggests that, like $\rm N_2H^+$,  $\rm DCO^+$ does not suffer 
significantly from depletion.
Regardless of depletion effects, however, it is evident 
that there is still enough optical
depth, excitation gradient, and relative gas motion for the  CS(3-2) 
and CS (2-1) line profiles to exhibit infall asymmetry.

The $\rm \delta V$ distribution of CS(3-2), with respect to the $\rm N_2H^+$, 
is found to be significantly skewed to the blue suggesting, again, that  CS(3-2) traces 
inward motions in the cores. Interestingly, the distribution is found to be statistically 
very similar to the $\rm \delta V$ distribution for CS(2-1) 
in spite of the higher infall velocities seen in CS(3-2).
This occurs because the higher infall speed for  CS(3-2) is not large enough to change 
the histogram of $\rm \delta V$ distribution. 
Altogether, this means that inward motions may occur over a substantial range of core densities,
and the lower density and higher density gas probed by CS(2-1) and (3-2) 
lines respectively move together, in a systematic rather than random fashion.
On the other hand, the $\rm \delta V$ distribution of DCO$^+$(2-1) 
(with respect to $\rm N_2H^+$) gives a rather symmetric $\rm \delta V$ distribution. 
We attribute the relative smaller incidence of line asymmetry in DCO$^+$(2-1) than in 
CS(3-2) and in CS(2-1) to lower optical depth of DCO$^+$(2-1) than in the CS lines.

\acknowledgments
This research was supported  by NASA Origins of Solar System Program, Grant NAGW-5626.
C.W.L. greatly acknowledges supports from the Basic Research Program 
(KOSEF R01-2003-000-10513-0) of the Korea Science and Engineering Foundation, and 
Strategic National R\&D Program (M1-0222-00-0005)
from Ministry of Science and Technology, Republic of Korea.  RP is also supported by the Natural
Sciences and Engineering Research Council of Canada.

\vfill\eject

\clearpage

%\small
%\input{tab1.tex}

%\input{tab2.tex}

%\input{tab3.tex}

%\clearpage

%\addtocounter{page}{6}
\begin{figure}
\begin{center}
{\bf FIGURE CAPTIONS}
\end{center}
\end{figure}

\begin{figure}
%\plotone{f1-1.ps} 
\noindent{\bf Fig. 1. ---} 
CS(3-2) and DCO$^+$  spectra. The dashed line indicates 
the position of the centroid velocity as derived by a Gaussian fit  
to $\rm N_2H^+~1-0$ from LMT99. 
Note that in the figures source names starting with ``DC'' are abbreviated 
with short ones; DC3382+164 with DC164, DC3388+165-3 with DC165-3, 
DC3388+165-6 with DC165-6, DC3392+161-3 with DC161-3, and DC3392+161-4 with DC161-4. 

\end{figure}

\begin{figure}
%\plotone{f2.ps} 
\noindent{\bf Fig. 2. --- } 
CS(3-2) and CS(2-1) spectra of three starless cores illustrating 
that CS(2-1) has broader wings than CS(3-2). 

\end{figure}

\begin{figure}
%\plotone{f3.ps} 
\noindent{\bf Fig. 3. --- } 
Histograms of the normalized velocity differences between CS(3-2) and 
$\rm N_2H^+~1-0$, and CS(2-1) and $\rm N_2H^+~1-0$.

\end{figure}

\begin{figure}
%\plotone{f4.ps} 
\noindent{\bf Fig. 4. --- } 
Correlation between the normalized velocity differences of CS(3-2) and CS(2-1) for all sources.
\end{figure}

\begin{figure}
%\plotone{f5.ps} 
\noindent{\bf Fig. 5. --- } 
Histogram of the normalized velocity difference  ($\rm  \delta V_{DCO^+}$ )
between $\rm DCO^+$(2-1) and $\rm N_2H^+$(1-0).
\end{figure}

\begin{figure}
%\plotone{f6.ps} 
\noindent{\bf Fig. 6. --- } 
Two-layer radiative transfer model fits to CS spectra in twelve infall candidates. 
Twelve sources for which the CS(3-2) or (2-1) profiles show infall asymmetry
were chosen because their spectra have
either double peaks or  a  blue peak with a clear red shoulder.  
These types of spectra allow
us to accurately determine the infall speeds with small uncertainties.
Note that the model was not fit to the CS(2-1) spectra for two sources
(L1355 and L1155C-1), even though they show infall asymmetry,
because they  show neither clear
double peaks nor a red shoulder in their spectra and so fits to  
their spectra would result in large errors in the derived infall speeds.
The CS(2-1) spectra in  two additional sources (L234E-S and L429-1)  
were not fit
to the model since they do not exhibit spectral infall features.
\end{figure}

\begin{figure}
%\plotone{f7.ps} 
\noindent{\bf Fig. 7. --- } 
Infall speeds derived from our two layer model fits to CS(2-1) and CS(3-2) profiles.
The error bars  for L1521F represent 1 sigma uncertainties of 
the distribution of infall speeds obtained from repeated fits of two-layer models 
to each twenty synthetic CS(2-1) and (3-2) spectra (as described in $\S$ 4.1).
\end{figure}

\begin{figure}
%\plotone{f8.ps} 
\noindent{\bf Fig. 8. --- }  
Reliability tests of our simple two layers model in deriving infall speeds in L1521F.
(a) Distribution of infall speeds derived from repeated fits of two-layer models
to CS(2-1) and (3-2) spectra for L1521F. The mean and 1 sigma uncertainty of the distributions
are $\rm 0.014\pm0.001~km~s^{-1}$ and $\rm 0.045\pm0.003~km~s^{-1}$ for the CS(2-1). 
Figures (b) and (c) compare the observed spectral line profiles to the model fits 
derived using both the best fit velocities for each transition (solid lines) and 
an arbitrary velocity (dashed lines).
Solid lines in (b) and (c) indicate the best $\chi^2$ fit to the
CS(2-1) and (3-2) spectra in L1521F, producing infall speeds of 
$0.014 \rm ~km~s^{-1}$ and an infall speed of $0.045 \rm ~km~s^{-1}$ for (2-1) and (3-2)
spectra, respectively. Dashed lines in
(b) and (c) show the spectra obtained if
we attempt to fit the CS(2-1) spectrum with the CS(3-2) velocity (and  
vice-versa), revealing that small  
changes in the infall velocity result in poor fits to the spectra as well as 
higher $\chi^2$ values. Here the $\chi^2$ values given in the figure are normalized 
with the number of degree of freedoms (107 for CS 2-1  and 78 for CS 3-2). 

\end{figure}

\begin{figure}
%\plotone{f9.ps} 
\noindent{\bf Fig. 9. --- }  
The same test for the optical depth as Fig. 8.
(a) Distribution of optical depths derived from repeated fits of two-layer models
to CS(2-1) and (3-2) spectra for L1521F. The 1 sigma uncertainty of the distributions
$\sim 0.035$ for the CS(2-1) and $\sim 0.03$ for the CS(3-2).
(b) and (c) are comparisons of fits of CS spectra using their best fit optical depths in L1521F 
with the fits of the lines using other arbitrary optical depths (here CS 3-2 optical depth for 
the 2-1 line and CS 2-1 optical depth for the 3-2 line).  
Solid lines in (b) and (c) indicate the best $\chi^2$ fit to the
CS(2-1) and (3-2) spectra in L1521F. Dashed lines in
(b) and (c) shows the spectra obtained if
we attempt to fit the CS(2-1) spectrum with the CS(3-2) optical depth (and  
vice-versa), again revealing that small  
changes in the optical depth cause worse fits to the spectra.

\end{figure}

\begin{figure}
%\plotone{f10.ps} 
\noindent{\bf Fig. 10. --- } 
The distribution of the sources in infall speed difference ($\rm V_{in,32} - V_{in,21}$)
versus optical depth difference ($\tau_{21} - \tau_{32}$) between two CS(3-2) and (2-1) lines.
Sources marked with open circles use the FCRAO CS(2-1) data  while sources with 
filled circles use the Haystack CS(2-1) data. 
Data points for the same source are joined by a line. 
\end{figure}

\begin{figure}
%\plotone{f11.ps} 
\noindent{\bf Fig. 11. --- } 
Integrated intensities of CS, $\rm DCO^+$, and $\rm N_2H^+$
for starless cores. 
The figure shows little correlation between the CS and $\rm N_2H^+$ intensities,
but a fairly good correlation between $\rm DCO^+$ and $\rm N_2H^+$ intensities,
implying ubiquitous depletion of CS but little depletion of $\rm DCO^+$.
\end{figure}

\end{document}

%\addtocounter{page}{}
\begin{deluxetable}{lcccccccccr}
%\tablewidth{50pc}
\small
\tablecaption{Line Asymmetry Colors of Starless Cores\tablenotemark{a}}
\tablehead{
        \colhead{Source}   &\colhead{}     & \colhead{From $\rm \delta V$} & \colhead{} &
        \colhead{} & \colhead{} & \colhead{From $\rm T_b/T_r$} &\colhead{}  &
        \colhead{rating }&\colhead{N} \nl
        \colhead{}   &\colhead{[$\rm CS$(2-1)} & \colhead{$\rm CS$(3-2)} & 
        \colhead{$\rm DCO^+$(2-1)]} & \colhead{[$\rm CS$(2-1)} & 
        \colhead{$\rm CS$(3-2)} & \colhead{$\rm DCO^+$(2-1)} & \colhead{$\rm N_2H^+$(1-0)]} &
        \colhead{index (C)} & \colhead{} 
        }
\startdata
L1333       &   0   & 0     &-1     &  \nodata&  \nodata&  \nodata&  \nodata& -1  &3\nl
L1355       &   1   & 1     & 0     &    1    &     1   &  \nodata&  \nodata&  4  &5\nl
L1498       &   1   & 1     & 0     &    1    &     1   &  \nodata&  \nodata&  4  &5\nl
L1399-1     &   0   & 0     & 0     &  \nodata&     1   &  \nodata&  \nodata&  1  &4\nl
L1521B-2    &   0   & 0     & 0     &  \nodata&  \nodata&  \nodata&  \nodata&  0  &3\nl
L1399-2     &   0   & 0     & 0     &  \nodata&  \nodata&  \nodata&  \nodata&  0  &3\nl
L1399-3     &   0   & 0     & 0     &  \nodata&  \nodata&  \nodata&  \nodata&  0  &3\nl
B217-2      &   0   & 0     & 0     &  \nodata&  \nodata&  \nodata&  \nodata&  0  &3\nl
L1521F      &   1   & 1     & 1     &    1    &    1    &  \nodata&    1    &  6  &6\nl
L1521-2     &   0   & 0     & 0     &  \nodata&  \nodata&  \nodata&  \nodata&  0  &3\nl
L1400A      &   0   & 1     & 0     &  \nodata&  \nodata&  \nodata&  \nodata&  1  &3\nl
L1524-4     &   1   & 0     & 0     &  \nodata&  \nodata&  \nodata&  \nodata&  1  &3\nl
L1551S-2    &   0   & 0     & 0     &  \nodata&  \nodata&  \nodata&  \nodata&  0  &3\nl
L1445       &\nodata&\nodata& 0     &    1    &    1    &  \nodata&  \nodata&  2  &3\nl
TMC2        &   1   & 1     & 0     &  \nodata&  \nodata&  \nodata&  \nodata&  2  &3\nl
B18-3       &   0   & 0     & 0     &    1    &  \nodata&  \nodata&  \nodata&  1  &4\nl
B18-5       &   0   & 0     &-1     &  \nodata&  \nodata&  \nodata&  \nodata& -1  &3\nl
CB22        &   0   & 0     & 0     &  \nodata&  \nodata&  \nodata&  \nodata&  0  &3\nl
TMC1        &   1   & 1     & 0     &  \nodata&  \nodata&  \nodata&  \nodata&  2  &3\nl
L1507A-1    &   0   & 0     & 0     &  \nodata&  \nodata&  \nodata&  \nodata&  0  &3\nl
CB23        &   1   & 0     & 0     &  \nodata&  \nodata&  \nodata&  \nodata&  1  &3\nl
L1517A      &   0   & 0     & 0     &  \nodata&  \nodata&  \nodata&  \nodata&  0  &3\nl
L1517B      &   0   &-1     &-1     &  \nodata&  \nodata&  \nodata&  \nodata& -2  &3\nl
L1512-1     &   0   & 0     & 0     &  \nodata&  \nodata&  \nodata&  \nodata&  0  &3\nl
L1544       &\nodata& 1     & 1     &    1    &    1    &  \nodata&    1    &  5  &5\nl
L1552       &   0   & 1     & 0     &  \nodata&    1    &  \nodata&  \nodata&  2  &4\nl
L1582A      &   0   & 0     & 0     &  \nodata&    1    &  \nodata&  \nodata&  1  &4\nl
L1622A-2    &   1   & 0     &-1     &  \nodata&  \nodata&  \nodata&  \nodata&  0  &3\nl
L1622A-1    &   0   & 0     & 0     &  \nodata&  \nodata&  \nodata&  \nodata&  0  &3\nl
CB45        &   0   & 0     &-1     &  \nodata&  \nodata&  \nodata&  \nodata& -1  &3\nl
DC3392+161-4&\nodata&\nodata&\nodata&  \nodata&   -1    &  \nodata&  \nodata& -1  &1\nl
L134A       &   0   & 0     &-1     &  \nodata&  \nodata&  \nodata&  \nodata& -1  &3\nl
L183        &   1   & 1     &-1     &    1    &    1    &   -1    &  \nodata&  2  &6\nl
L1696A      &   1   & 0     & 0     &  \nodata&    1    &  \nodata&  \nodata&  2  &3\nl
L1696B      &   0   & 0     & 1     &  \nodata&  \nodata&  \nodata&  \nodata&  1  &3\nl
L1704-1     &   0   & 0     & 0     &  \nodata&  \nodata&  \nodata&  \nodata&  0  &3\nl
L1709B-2    &   0   & 0     & 1     &  \nodata&  \nodata&  \nodata&  \nodata&  1  &3\nl
L1689B      &   0   & 0     & 0     &    1    &    1    &  \nodata&  \nodata&  2  &5\nl
L1782-2     &   0   & 0     & 0     &  \nodata&  \nodata&  \nodata&  \nodata&  0  &3\nl
L158        &   1   & 1     & 1     &    1    &  \nodata&  \nodata&  \nodata&  4  &4\nl
L204C-2     &   0   & 1     & 0     &  \nodata&  \nodata&  \nodata&  \nodata&  1  &3\nl
L234E-S     &   1   & 1     &-1     &  \nodata&    1    &  \nodata&  \nodata&  2  &4\nl
L63         &  -1   &\nodata& 1     &  \nodata&  \nodata&  \nodata&  \nodata&  0  &2\nl
L462-1      &   0   & 0     & 0     &  \nodata&  \nodata&  \nodata&  \nodata&  0  &3\nl
L462-2      &   0   & 0     & 0     &  \nodata&  \nodata&  \nodata&  \nodata&  0  &3\nl
L492        &   1   & 1     & 1     &  \nodata&  \nodata&    1    &    0    &  4  &5\nl
CB130-1     &\nodata& 0     & 0     &  \nodata&  \nodata&  \nodata&  \nodata&  0  &2\nl
CB130-3     &\nodata& 0     & 0     &  \nodata&  \nodata&  \nodata&  \nodata&  0  &2\nl
L429-1      &\nodata& 1     & 0     &    0    &    1    &  \nodata&    0    &  2  &5\nl
L648-1      &   0   &\nodata& 0     &  \nodata&  \nodata&  \nodata&  \nodata&  0  &2\nl
L673-7      &   0   & 0     & 0     &  \nodata&  \nodata&  \nodata&  \nodata&  0  &3\nl
L694-2      &   1   & 1     & 1     &    1    &    1    &    1    &  \nodata&  6  &6\nl
L1041-2     &   0   & 0     & 0     &  \nodata&  \nodata&  \nodata&  \nodata&  0  &3\nl
L922-1      &   0   & 0     & 0     &  \nodata&  \nodata&  \nodata&  \nodata&  0  &3\nl
L922-2      &   1   & 0     & 0     &  \nodata&  \nodata&  \nodata&  \nodata&  1  &3\nl
L1049-1     &  -1   & 0     & 0     &  \nodata&  \nodata&  \nodata&  \nodata& -1  &3\nl
L1049-2     &   0   & 0     & 0     &  \nodata&  \nodata&  \nodata&  \nodata&  0  &3\nl
L1148       &   0   & 0     & 0     &  \nodata&  \nodata&  \nodata&  \nodata&  0  &3\nl
L1155C-2    &   1   & 0     & 0     &  \nodata&  \nodata&  \nodata&  \nodata&  1  &3\nl
L1155C-1    &   1   & 1     & 0     &    1    &    1    &  \nodata&  \nodata&  4  &5\nl
L1155E      &   0   & 0     & 0     &  \nodata&  \nodata&  \nodata&  \nodata&  0  &3\nl
L1082C-2    &   0   & 0     & 0     &  \nodata&  \nodata&  \nodata&  \nodata&  0  &3\nl
L981-1      &   0   & 0     &-1     &  \nodata&  \nodata&  \nodata&  \nodata& -1  &3\nl
L1063       &   0   & 0     & 0     &  \nodata&    1    &  \nodata&  \nodata&  1  &4\nl
L944-2      &   0   & 0     & 0     &  \nodata&  \nodata&  \nodata&  \nodata&  0  &3\nl
L1251A-2    &   0   & 0     & 0     &  \nodata&  \nodata&  \nodata&  \nodata&  0  &3\nl
L1197       &   0   & 0     & 0     &  \nodata&  \nodata&    1    &  \nodata&  1  &4\nl
L1251       &   0   & 0     & 0     &  \nodata&  \nodata&  \nodata&  \nodata&  0  &3\nl
L1234       &   1   & 1     & 0     &  \nodata&  \nodata&  \nodata&  \nodata&  2  &3\nl
CB246-2     &  -1   &-1     & 0     &   -1    &   -1    &  \nodata&  \nodata& -4  &5\nl
\enddata								          
\tablenotetext{a}{The table gives a summary of line asymmetry colors of starless cores 
with CS (2-1) \& $\rm N_2H^+$ (1-0) observations (LMT99), and CS(3-2) \& $\rm DCO^+$(2-1)
observations (this study).
The column 1 lists the name of source, and 
the columns 2 to 4 illustrate a diagnostics from the normalized velocity differences 
($\delta V$) of CS(2-1), CS(3-2) and DCO$^+$(2-1) with respect to $\rm N_2H^+$ velocities from LMT99.
The numeric numbers `1' or `-1' are given for sources with blue or red
asymmetry of $\vert \delta V\vert \ge 5~\sigma_{\delta V}$, 
and `0' for sources with neither asymmetry of 
$\vert \delta V\vert \le 5~\sigma_{\delta V}$.
The columns 5 to 8 shows other diagnostics for the line asymmetry colors using the 
ratio of intensities of blue component ($\rm T_b$) and red component ($\rm T_r$)
obtained from the Gaussian fit for double-peaked spectra of CS(2-1), CS(3-2), DCO$^+$(2-1),
and $\rm N_2H^+$(1-0) main component.
Here `1', `0', or `-1' are designated for the sources 
of $\rm T_b/T_r > 1+\sigma$, $\rm 1-\sigma \le T_b/T_r \le
1+\sigma$, or $\rm T_b/T_r < 1-\sigma$, respectively, where $\sigma$ is 
one propagation error at each value of the ratio. 
The column 9 gives the algebraic sum of the individual color codes 
of each source, a rating index C. The N in the last column is  
the total number of spectra considered for color codes of each core.}
 
\end{deluxetable}

%\addtocounter{page}{}
\begin{deluxetable}{lccccrrrrc}
%\tablewidth{50pc}
\small
\tablecaption{ Starless Cores Observed in CS $(3-2)$ and/or $\rm DCO^+~(2-1)$ lines\tablenotemark{a}}
\tablehead{
        \colhead{Source}   &\colhead{RA (1950.0)}     & \colhead{DEC (1950.0)}          &
        \colhead{$\rm T_A^*(CS)$} &\colhead{$\rm T_A^*(DCO^+)$} & \colhead{$\rm V_{CS}$} &
        \colhead{$\rm V_{DCO^+}$} &\colhead{$\rm \delta V_{CS}$}   &
        \colhead{$\rm \delta V_{DCO^+}$} & \colhead{Source} \nl
        \colhead{}     & \colhead{$^h~~^m~~^s~$}  &\colhead{~~~~$^{\circ}~~~'~~~''$}&
        \colhead{(K)}           & \colhead{(K)}  & \colhead{(km~s$^{-1}$)}         &
        \colhead{(km~s$^{-1}$)}           &\colhead{}   & \colhead{} & \colhead{Selection}
        }
\startdata
L1333         &02~21~10.2 & 75~13~32&     0.36 &    1.13 &  3.85$\pm$0.01&  3.91$\pm$0.00&  0.04$\pm$0.03&  0.19$\pm$0.02&1 \nl
L1355         &02~48~41.4 & 68~43~36&     0.90 &    0.22 & -3.86$\pm$0.01& -3.71$\pm$0.01& -0.56$\pm$0.08&  0.02$\pm$0.05&1 \nl
B5-1          &03~44~26.4 & 32~46~45&     0.56 &    0.21 &  9.58$\pm$0.01&  9.65$\pm$0.01& \nodata       & \nodata       &1 \nl
L1498         &04~07~50.0 & 25~02~05&     0.77 &    0.41 &  7.73$\pm$0.00&  7.89$\pm$0.01& -0.56$\pm$0.05& -0.16$\pm$0.05&1 \nl
L1399-1       &04~20~00.5 & 55~01~00&     0.25 &    0.27 &  3.23$\pm$0.01&  3.32$\pm$0.01& -0.08$\pm$0.06&  0.08$\pm$0.06&1 \nl
L1521B-2      &04~20~32.8 & 26~33~11&     0.51 &    0.24 &  6.44$\pm$0.00&  6.45$\pm$0.01& -0.13$\pm$0.12& -0.13$\pm$0.13&1 \nl
L1399-2       &04~20~58.3 & 54~52~48&     0.23 &    0.29 &  3.57$\pm$0.03&  3.42$\pm$0.01&  0.31$\pm$0.07& -0.02$\pm$0.04&1 \nl
L1399-3       &04~21~11.8 & 54~49~59&     0.17 &    0.16 &  3.62$\pm$0.03&  3.62$\pm$0.01& -0.04$\pm$0.08& -0.01$\pm$0.07&1 \nl
B217-2        &04~25~04.5 & 26~14~16&     1.27 &    1.10 &  6.84$\pm$0.00&  6.90$\pm$0.00& -0.10$\pm$0.02&  0.04$\pm$0.02&1 \nl
L1400A-2      &04~25~05.8 & 54~39~40& $\le0.06$& \nodata &  \nodata      &  \nodata      & \nodata       & \nodata       &1 \nl
L1521F        &04~25~35.0 & 26~45~00&     1.07 &    1.17 &  6.27$\pm$0.00&  6.42$\pm$0.00& -0.72$\pm$0.02& -0.29$\pm$0.02&1 \nl
L1521-2       &04~26~26.8 & 26~53~27&     0.48 &    0.09 &  6.64$\pm$0.01&  6.49$\pm$0.01&  0.44$\pm$0.11& -0.19$\pm$0.09&1 \nl
%L1400K       &04~26~45.3 & 54~44~37&     0.39 &    0.28 &  3.09$\pm$0.01&  3.11$\pm$0.01&               &               &  \nl
L1400A        &04~26~50.6 & 54~46~08&     0.36 &    0.27 &  3.23$\pm$0.01&  3.31$\pm$0.01& -0.44$\pm$0.07& -0.10$\pm$0.05&1 \nl
L1524-4       &04~27~04.1 & 24~18~47&     0.72 &    0.27 &  6.24$\pm$0.01&  6.28$\pm$0.00&  0.02$\pm$0.03&  0.13$\pm$0.03&1 \nl
L1551S-2      &04~28~03.5 & 18~09~09&     0.19 &    0.07 &  6.59$\pm$0.01&  6.62$\pm$0.03& -0.31$\pm$0.07& -0.25$\pm$0.10&1 \nl
L1445         &04~28~28.8 & 46~31~01&     0.36 &    0.07 &  5.22$\pm$0.01&  5.32$\pm$0.01& \nodata       & \nodata       &1 \nl
TMC2          &04~29~46.9 & 24~18~54&     0.76 &    0.87 &  6.10$\pm$0.01&  6.24$\pm$0.00& -0.48$\pm$0.06& -0.04$\pm$0.03&1 \nl
B18-3         &04~31~04.6 & 24~03~32&     0.87 &    0.29 &  6.85$\pm$0.01&  6.84$\pm$0.00& -0.43$\pm$0.16& -0.49$\pm$0.17&1 \nl
B18-5         &04~32~49.7 & 24~03~15&     0.62 &    0.68 &  6.17$\pm$0.01&  6.39$\pm$0.00& -0.26$\pm$0.06&  0.33$\pm$0.06&1 \nl
CB22          &04~37~30.2 & 29~47~13&     0.53 &    0.34 &  5.99$\pm$0.00&  6.03$\pm$0.01& -0.04$\pm$0.04&  0.07$\pm$0.05&1 \nl
TMC1          &04~38~29.0 & 25~39~01&     2.10 &    0.75 &  5.77$\pm$0.00&  5.92$\pm$0.00& -0.30$\pm$0.02&  0.05$\pm$0.02&1 \nl
L1507A-1      &04~39~29.0 & 29~38~07&     0.60 &    0.63 &  6.21$\pm$0.01&  6.23$\pm$0.01& -0.08$\pm$0.05& -0.02$\pm$0.04&1 \nl
CB23          &04~40~22.0 & 29~33~36&     0.54 &    0.35 &  6.04$\pm$0.00&  6.07$\pm$0.01& -0.14$\pm$0.04& -0.07$\pm$0.04&1 \nl
L1517C-1      &04~51~35.9 & 30~30~50&     0.67 &    0.09 &  5.55$\pm$0.01&  5.57$\pm$0.01& \nodata       & \nodata       &1 \nl
L1517A        &04~51~58.7 & 30~28~04&     0.59 &    0.20 &  5.80$\pm$0.00&  5.81$\pm$0.01& -0.02$\pm$0.08& -0.01$\pm$0.09&1 \nl
L1517B        &04~52~07.2 & 30~33~18&     0.54 &    0.55 &  5.88$\pm$0.00&  5.87$\pm$0.00&  0.26$\pm$0.03&  0.23$\pm$0.03&1 \nl
L1512-1       &05~00~54.4 & 32~39~00&     0.57 &    0.61 &  7.15$\pm$0.00&  7.16$\pm$0.00& -0.01$\pm$0.04& -0.01$\pm$0.03&1 \nl
L1544         &05~01~12.5 & 25~06~40&     1.14 &    0.36 &  7.07$\pm$0.01&  7.12$\pm$0.02& -0.27$\pm$0.03& -0.12$\pm$0.02&1 \nl
L1544-2       &05~01~39.1 & 25~07~30&  \nodata &$\le0.05$&  \nodata      &  \nodata      & \nodata       & \nodata       &1 \nl
CB28          &05~03~51.2 &-04~00~20&     1.04 &    0.26 &  8.80$\pm$0.01&  8.97$\pm$0.01& \nodata       & \nodata       &1 \nl
L1552         &05~14~31.6 & 26~02~07&     1.23 &    1.13 &  7.59$\pm$0.01&  7.68$\pm$0.00& -0.37$\pm$0.06&  0.00$\pm$0.04&1 \nl
L1582A        &05~29~15.0 & 12~28~57&     0.72 &    0.58 & 10.08$\pm$0.01& 10.28$\pm$0.01& -0.31$\pm$0.07&  0.06$\pm$0.05&1 \nl
L1622A-2      &05~52~02.9 & 01~53~15&     0.88 &    0.85 &  1.11$\pm$0.01&  1.19$\pm$0.00& -0.08$\pm$0.03&  0.17$\pm$0.02&1 \nl
L1622A-1      &05~52~17.1 & 01~56~55&     1.03 &    0.93 &  1.21$\pm$0.01&  1.21$\pm$0.00&  0.09$\pm$0.03&  0.07$\pm$0.02&1 \nl
L1622B        &05~52~17.5 & 01~45~11&     0.23 &$\le0.05$&  0.51$\pm$0.01&  \nodata      & \nodata       & \nodata       &2 \nl
L1578-2       &06~05~40.4 & 18~12~00&     0.31 &$\le0.10$& -0.42$\pm$0.01&  \nodata      & \nodata       & \nodata       &2 \nl
%L1578-1      &06~05~41.6 & 18~06~54&     0.30 &    0.05 & -0.35$\pm$0.01& -0.35$\pm$0.02& \nodata       & \nodata       &2 \nl
CB45          &06~06~02.0 & 17~50~48&     0.27 &    0.56 &  0.42$\pm$0.01&  0.56$\pm$0.01&  0.07$\pm$0.04&  0.31$\pm$0.04&1 \nl
V13-2         &08~06~54.9 &-35~54~21&     0.83 &    0.06 &  5.98$\pm$0.01&  6.08$\pm$0.04& \nodata       & \nodata       &2 \nl
V13-3         &08~07~01.0 &-35~48~22&     0.82 &    0.07 &  6.36$\pm$0.02&  6.32$\pm$0.03& \nodata       & \nodata       &2 \nl
V13-1         &08~07~09.0 &-35~52~18&     1.31 &    0.31 &  6.24$\pm$0.01&  6.35$\pm$0.02& \nodata       & \nodata       &2 \nl
DC3382+164    &15~37~23.9 &-34~30~40&     0.88 &    0.40 &  4.70$\pm$0.01&  4.67$\pm$0.01& \nodata       & \nodata       &2 \nl
DC3388+165-2  &15~39~08.8 &-33~41~27&  \nodata &$\le0.08$&  \nodata      &  \nodata      & \nodata       & \nodata       &2 \nl
DC3388+165-3  &15~39~29.2 &-33~42~30&     0.38 &    0.07 &  4.46$\pm$0.01&  4.61$\pm$0.03& \nodata       & \nodata       &2 \nl
DC3388+165-6  &15~40~06.5 &-34~04~01&     0.35 &    0.24 &  5.30$\pm$0.01&  5.34$\pm$0.02& \nodata       & \nodata       &2 \nl
DC3392+161-2  &15~42~01.1 &-34~03~59&     0.55 &$\le0.09$&  5.36$\pm$0.02&  \nodata      & \nodata       & \nodata       &2 \nl
DC3392+161-3  &15~42~03.2 &-34~11~21&     0.44 &    0.26 &  5.05$\pm$0.01&  5.07$\pm$0.02& \nodata       & \nodata       &2 \nl
DC3392+161-4  &15~42~16.9 &-34~15~19&     0.55 &    0.21 &  4.44$\pm$0.01&  4.38$\pm$0.02& \nodata       & \nodata       &2 \nl
L134A         &15~50~58.1 &-04~26~36&     0.40 &    0.62 &  2.78$\pm$0.01&  2.78$\pm$0.01&  0.16$\pm$0.04&  0.15$\pm$0.03&1 \nl
L183          &15~51~30.0 &-02~42~51&     0.43 &    1.28 &  2.35$\pm$0.01&  2.69$\pm$0.00& -0.59$\pm$0.05&  0.80$\pm$0.02&1 \nl
%L183         &15~51~35.7 &-02~40~54&     0.61 &    0.99 &  2.38$\pm$0.01&  2.49$\pm$0.00& -0.28$\pm$0.02& -0.28$\pm$0.02&1 \nl
L1696A        &16~25~30.0 &-24~11~42&     1.47 &    2.81 &  3.39$\pm$0.00&  3.37$\pm$0.00& -0.06$\pm$0.03& -0.17$\pm$0.04&1 \nl
L1696B        &16~25~57.8 &-24~14~09&     0.97 &    2.19 &  3.28$\pm$0.00&  3.25$\pm$0.00& -0.19$\pm$0.05& -0.32$\pm$0.06&1 \nl
L1704-1       &16~27~49.8 &-23~35~49&     0.69 &    1.25 &  2.72$\pm$0.01&  2.77$\pm$0.00&  0.07$\pm$0.05&  0.25$\pm$0.07&1 \nl
L1709A        &16~27~50.0 &-23~34~37&     0.66 &    1.08 &  2.77$\pm$0.01&  2.70$\pm$0.01& \nodata       & \nodata       &1 \nl
L1757-1       &16~29~01.9 &-19~35~24&     0.10 & \nodata &  1.48$\pm$0.00&  \nodata      & \nodata       & \nodata       &1 \nl
L1709B-2      &16~29~49.3 &-23~46~22&     1.20 &    1.23 &  2.57$\pm$0.00&  2.52$\pm$0.00& -0.09$\pm$0.03& -0.25$\pm$0.03&1 \nl
L1689B        &16~31~43.6 &-24~31~40&     1.58 &    0.88 &  3.39$\pm$0.00&  3.53$\pm$0.01& -0.58$\pm$0.12&  0.03$\pm$0.14&1 \nl
DC3460+78     &16~33~34.7 &-35~30~27&     0.51 &    0.10 &  3.26$\pm$0.01&  3.29$\pm$0.02& \nodata       & \nodata       &2 \nl
DC3464+79     &16~34~13.4 &-35~08~26&     0.43 &    0.31 &  3.59$\pm$0.01&  3.60$\pm$0.01& \nodata       & \nodata       &2 \nl
L1782-2       &16~39~36.4 &-19~38~00&     0.66 &    0.38 &  5.38$\pm$0.01&  5.43$\pm$0.01& -0.14$\pm$0.06&  0.02$\pm$0.07&1 \nl
L158          &16~44~37.7 &-13~54~03&     0.39 &    1.49 &  3.82$\pm$0.01&  3.83$\pm$0.00& -0.38$\pm$0.05& -0.35$\pm$0.04&1 \nl
L204C         &16~44~44.2 &-12~04~27&     0.52 &$\le0.24$&  4.31$\pm$0.01&  \nodata      & \nodata       & \nodata       &1 \nl
L204C-2       &16~44~58.5 &-12~18~04&     0.97 &    0.39 &  4.12$\pm$0.01&  4.21$\pm$0.01& -0.33$\pm$0.06& -0.15$\pm$0.06&1 \nl
L234E-S       &16~45~22.6 &-10~52~10&     1.18 &    1.17 &  2.93$\pm$0.00&  3.29$\pm$0.00& -0.55$\pm$0.05&  0.48$\pm$0.05&1 \nl
L63           &16~47~21.0 &-18~00~10&     0.50 &    1.52 &  5.96$\pm$0.01&  5.72$\pm$0.01& \nodata       & -0.39$\pm$0.05&1 \nl
L462-1        &18~04~51.9 &-04~43~57&     0.57 &    0.24 &  7.90$\pm$0.00&  7.92$\pm$0.01& -0.06$\pm$0.04& -0.02$\pm$0.05&1 \nl
L462-2        &18~04~56.9 &-04~40~57&     0.52 &    0.40 &  7.94$\pm$0.01&  7.95$\pm$0.01&  0.01$\pm$0.04&  0.02$\pm$0.03&1 \nl
L492          &18~13~11.6 &-03~47~17&     1.63 &    0.42 &  7.64$\pm$0.01&  7.68$\pm$0.01& -0.23$\pm$0.02& -0.16$\pm$0.02&1 \nl
CB130-1       &18~13~39.2 &-02~34~06&     0.49 &    0.51 &  7.63$\pm$0.01&  7.65$\pm$0.01& -0.09$\pm$0.07& -0.05$\pm$0.07&1 \nl
CB130-3       &18~13~41.5 &-02~17~47&     0.73 &    0.61 &  7.25$\pm$0.01&  7.22$\pm$0.01&  0.18$\pm$0.04&  0.01$\pm$0.04&1 \nl
L429-1        &18~14~28.2 &-08~14~40&     0.94 &    0.71 &  6.69$\pm$0.00&  6.87$\pm$0.01& -0.64$\pm$0.07& -0.06$\pm$0.04&1 \nl
L429-2        &18~14~32.8 &-08~19~02&     0.14 &$\le0.07$&  6.59$\pm$0.02&  \nodata      & \nodata       & \nodata       &1 \nl
L451          &18~19~12.3 &-07~14~11&     0.14 & \nodata &  8.65$\pm$0.03&  \nodata      & \nodata       & \nodata       &1 \nl
L648-1        &18~37~09.7 & 12~38~37&     0.47 &    0.50 &  4.14$\pm$0.01&  4.13$\pm$0.00& \nodata       &  0.09$\pm$0.06&1 \nl
L673-7        &19~19~14.3 & 11~15~30&     0.56 &    0.68 &  7.09$\pm$0.01&  7.15$\pm$0.00& -0.12$\pm$0.03& -0.02$\pm$0.02&1 \nl
L694-2        &19~38~42.4 & 10~49~59&     0.63 &    1.64 &  9.44$\pm$0.00&  9.50$\pm$0.00& -0.60$\pm$0.02& -0.42$\pm$0.01&1 \nl
L1041-2       &20~36~05.9 & 57~38~48&     0.36 &    0.09 & -1.76$\pm$0.01& -1.66$\pm$0.02& -0.06$\pm$0.07&  0.19$\pm$0.09&1 \nl
L922-1        &20~36~46.5 & 44~21~24&     0.47 &    0.41 &  6.61$\pm$0.01&  6.64$\pm$0.01& -0.09$\pm$0.04& -0.01$\pm$0.05&1 \nl
L922-2        &20~37~04.6 & 44~25~20&     0.59 &    0.38 &  6.51$\pm$0.01&  6.52$\pm$0.01& -0.20$\pm$0.05& -0.20$\pm$0.08&1 \nl
L1049-1       &20~39~50.8 & 57~16~45&     0.43 &    0.29 &  0.28$\pm$0.02&  0.23$\pm$0.01&  0.23$\pm$0.05&  0.12$\pm$0.04&1 \nl
L1049-2       &20~40~16.8 & 57~22~57&     0.28 &    0.43 &  0.10$\pm$0.01&  0.10$\pm$0.01& -0.06$\pm$0.05& -0.08$\pm$0.05&1 \nl
L1148         &20~40~38.7 & 67~09~48&     0.75 &    0.17 &  2.56$\pm$0.00&  2.62$\pm$0.01& -0.17$\pm$0.04&  0.02$\pm$0.07&1 \nl
L1155C-2      &20~42~52.1 & 67~39~36&     0.41 &    0.32 &  1.35$\pm$0.01&  1.41$\pm$0.01& -0.17$\pm$0.04&  0.03$\pm$0.04&1 \nl
L1155C-1      &20~43~00.0 & 67~41~47&     0.42 &    0.37 &  2.58$\pm$0.01&  2.72$\pm$0.01& -0.31$\pm$0.03&  0.01$\pm$0.03&1 \nl
L1155E        &20~43~46.3 & 67~31~02&     0.20 &    0.06 &  2.46$\pm$0.01&  2.40$\pm$0.04&  0.01$\pm$0.09& -0.10$\pm$0.11&1 \nl
L1082C-2      &20~48~38.9 & 60~04~25&     0.55 &    0.59 & -2.42$\pm$0.01& -2.47$\pm$0.00&  0.29$\pm$0.06&  0.11$\pm$0.04&1 \nl
L981-1        &20~58~36.3 & 50~08~19&     0.32 &    0.14 & -0.12$\pm$0.01&  0.12$\pm$0.01& -0.03$\pm$0.05&  0.43$\pm$0.05&1 \nl
L1063         &21~06~44.6 & 56~06~09&     0.58 &    0.30 & -0.92$\pm$0.01& -0.81$\pm$0.00& -0.36$\pm$0.09& -0.07$\pm$0.07&1 \nl
L944-2        &21~15~54.0 & 43~05~41&     0.30 &    0.44 &  5.30$\pm$0.02&  5.42$\pm$0.00& -0.25$\pm$0.13&  0.10$\pm$0.12&1 \nl
L1251A-2      &22~26~46.5 & 74~58~10&     0.27 &    0.68 & -3.94$\pm$0.01& -3.90$\pm$0.01& -0.03$\pm$0.04&  0.05$\pm$0.03&1 \nl
L1197         &22~35~08.0 & 58~41~46&     0.40 &    0.18 & -3.16$\pm$0.01& -3.19$\pm$0.01& -0.12$\pm$0.05& -0.21$\pm$0.05&1 \nl
L1251         &22~36~26.0 & 74~54~46&     0.32 &    0.34 & -3.56$\pm$0.02& -3.50$\pm$0.01& -0.07$\pm$0.05&  0.03$\pm$0.04&1 \nl
%L1251E-1     &22~38~36.4 & 74~55~50&     0.97 &    1.60 & -4.60$\pm$0.01& -4.63$\pm$0.00&  0.04$\pm$0.01&  0.04$\pm$0.01&1 \nl
L1234         &23~15~47.5 & 62~10~15&     0.52 &    0.49 & -5.16$\pm$0.01& -5.04$\pm$0.01& -0.40$\pm$0.07& -0.10$\pm$0.06&1 \nl
CB246-2       &23~54~12.0 & 58~17~47&     0.60 &    0.65 & -0.68$\pm$0.01& -0.83$\pm$0.01&  0.43$\pm$0.06& -0.10$\pm$0.04&1 \nl
L1271-1       &23~57~32.6 & 67~05~07&     0.14 &$\le0.06$&-15.04$\pm$0.03&  \nodata      &  \nodata      &  \nodata      &1 \nl
\enddata								          
\tablenotetext{a}{The first column indicates the name of
the observed region, and the second and third columns 
list the central coordinates in B1950 epoch. The 4th and 5th column give 
antenna temperatures of CS $(3-2)$ and $\rm DCO^+ ~(2-1)$ spectra.
For undetected lines three sigma values of the observed lines are given.
The velocities of the detected lines listed in 6th and 7th column were obtained
by Gaussian fits of the lines. 
All designated errors  are 1 $\sigma$ values from the fits.
The 8th and 9th column list the normalized velocity differences ($\delta V$) 
of CS(3-2) and DCO$^+$ with respect to $\rm N_2H^+$ velocities from LMT99.
For example, $\delta V_{CS32}$ is defined as
$ \delta V_{CS32} = {{V_{CS32}-V_{N_2H^+}} \over {\Delta V_{N_2H^+}}}$.
The $1\sigma$ propagation errors are given at each value of $\delta V$.
The last column indicates the selection criterion for each source:
1 -- selected from LMT99, and 2 -- selected from LM99 catalog.
L1271-1 source has two more velocity components of $\rm V_{LSR}=-12.87,~-6.96~km~s^{-1}$. 
}
 
\end{deluxetable}

%\end{document}
% TABLE2.TEX -- AASTeX sample table 1.
 
\special{landscape}
\def\etal{{\it et al.~}}
\setlength{\textwidth}{15cm}
\setlength{\textheight}{20cm}
%\begin{document}
%\addtocounter{page}{}
\begin{deluxetable}{lcccccccccr}
%\tablewidth{50pc}
\small
\tablecaption{Line Asymmetry Colors of Starless Cores\tablenotemark{a}}
\tablehead{
        \colhead{Source}   &\colhead{}     & \colhead{From $\rm \delta V$} & \colhead{} &
        \colhead{} & \colhead{} & \colhead{From $\rm T_b/T_r$} &\colhead{}  &
        \colhead{rating }&\colhead{N} \nl
        \colhead{}   &\colhead{[$\rm CS$(2-1)} & \colhead{$\rm CS$(3-2)} & 
        \colhead{$\rm DCO^+$(2-1)]} & \colhead{[$\rm CS$(2-1)} & 
        \colhead{$\rm CS$(3-2)} & \colhead{$\rm DCO^+$(2-1)} & \colhead{$\rm N_2H^+$(1-0)]} &
        \colhead{index (C)} & \colhead{} 
        }
\startdata
L1333       &   0   & 0     &-1     &  \nodata&  \nodata&  \nodata&  \nodata& -1  &3\nl
L1355       &   1   & 1     & 0     &    1    &     1   &  \nodata&  \nodata&  4  &5\nl
L1498       &   1   & 1     & 0     &    1    &     1   &  \nodata&  \nodata&  4  &5\nl
L1399-1     &   0   & 0     & 0     &  \nodata&     1   &  \nodata&  \nodata&  1  &4\nl
L1521B-2    &   0   & 0     & 0     &  \nodata&  \nodata&  \nodata&  \nodata&  0  &3\nl
L1399-2     &   0   & 0     & 0     &  \nodata&  \nodata&  \nodata&  \nodata&  0  &3\nl
L1399-3     &   0   & 0     & 0     &  \nodata&  \nodata&  \nodata&  \nodata&  0  &3\nl
B217-2      &   0   & 0     & 0     &  \nodata&  \nodata&  \nodata&  \nodata&  0  &3\nl
L1521F      &   1   & 1     & 1     &    1    &    1    &  \nodata&    1    &  6  &6\nl
L1521-2     &   0   & 0     & 0     &  \nodata&  \nodata&  \nodata&  \nodata&  0  &3\nl
L1400A      &   0   & 1     & 0     &  \nodata&  \nodata&  \nodata&  \nodata&  1  &3\nl
L1524-4     &   1   & 0     & 0     &  \nodata&  \nodata&  \nodata&  \nodata&  1  &3\nl
L1551S-2    &   0   & 0     & 0     &  \nodata&  \nodata&  \nodata&  \nodata&  0  &3\nl
L1445       &\nodata&\nodata& 0     &    1    &    1    &  \nodata&  \nodata&  2  &3\nl
TMC2        &   1   & 1     & 0     &  \nodata&  \nodata&  \nodata&  \nodata&  2  &3\nl
B18-3       &   0   & 0     & 0     &    1    &  \nodata&  \nodata&  \nodata&  1  &4\nl
B18-5       &   0   & 0     &-1     &  \nodata&  \nodata&  \nodata&  \nodata& -1  &3\nl
CB22        &   0   & 0     & 0     &  \nodata&  \nodata&  \nodata&  \nodata&  0  &3\nl
TMC1        &   1   & 1     & 0     &  \nodata&  \nodata&  \nodata&  \nodata&  2  &3\nl
L1507A-1    &   0   & 0     & 0     &  \nodata&  \nodata&  \nodata&  \nodata&  0  &3\nl
CB23        &   1   & 0     & 0     &  \nodata&  \nodata&  \nodata&  \nodata&  1  &3\nl
L1517A      &   0   & 0     & 0     &  \nodata&  \nodata&  \nodata&  \nodata&  0  &3\nl
L1517B      &   0   &-1     &-1     &  \nodata&  \nodata&  \nodata&  \nodata& -2  &3\nl
L1512-1     &   0   & 0     & 0     &  \nodata&  \nodata&  \nodata&  \nodata&  0  &3\nl
L1544       &\nodata& 1     & 1     &    1    &    1    &  \nodata&    1    &  5  &5\nl
L1552       &   0   & 1     & 0     &  \nodata&    1    &  \nodata&  \nodata&  2  &4\nl
L1582A      &   0   & 0     & 0     &  \nodata&    1    &  \nodata&  \nodata&  1  &4\nl
L1622A-2    &   1   & 0     &-1     &  \nodata&  \nodata&  \nodata&  \nodata&  0  &3\nl
L1622A-1    &   0   & 0     & 0     &  \nodata&  \nodata&  \nodata&  \nodata&  0  &3\nl
CB45        &   0   & 0     &-1     &  \nodata&  \nodata&  \nodata&  \nodata& -1  &3\nl
DC3392+161-4&\nodata&\nodata&\nodata&  \nodata&   -1    &  \nodata&  \nodata& -1  &1\nl
L134A       &   0   & 0     &-1     &  \nodata&  \nodata&  \nodata&  \nodata& -1  &3\nl
L183        &   1   & 1     &-1     &    1    &    1    &   -1    &  \nodata&  2  &6\nl
L1696A      &   1   & 0     & 0     &  \nodata&    1    &  \nodata&  \nodata&  2  &3\nl
L1696B      &   0   & 0     & 1     &  \nodata&  \nodata&  \nodata&  \nodata&  1  &3\nl
L1704-1     &   0   & 0     & 0     &  \nodata&  \nodata&  \nodata&  \nodata&  0  &3\nl
L1709B-2    &   0   & 0     & 1     &  \nodata&  \nodata&  \nodata&  \nodata&  1  &3\nl
L1689B      &   0   & 0     & 0     &    1    &    1    &  \nodata&  \nodata&  2  &5\nl
L1782-2     &   0   & 0     & 0     &  \nodata&  \nodata&  \nodata&  \nodata&  0  &3\nl
L158        &   1   & 1     & 1     &    1    &  \nodata&  \nodata&  \nodata&  4  &4\nl
L204C-2     &   0   & 1     & 0     &  \nodata&  \nodata&  \nodata&  \nodata&  1  &3\nl
L234E-S     &   1   & 1     &-1     &  \nodata&    1    &  \nodata&  \nodata&  2  &4\nl
L63         &  -1   &\nodata& 1     &  \nodata&  \nodata&  \nodata&  \nodata&  0  &2\nl
L462-1      &   0   & 0     & 0     &  \nodata&  \nodata&  \nodata&  \nodata&  0  &3\nl
L462-2      &   0   & 0     & 0     &  \nodata&  \nodata&  \nodata&  \nodata&  0  &3\nl
L492        &   1   & 1     & 1     &  \nodata&  \nodata&    1    &    0    &  4  &5\nl
CB130-1     &\nodata& 0     & 0     &  \nodata&  \nodata&  \nodata&  \nodata&  0  &2\nl
CB130-3     &\nodata& 0     & 0     &  \nodata&  \nodata&  \nodata&  \nodata&  0  &2\nl
L429-1      &\nodata& 1     & 0     &    0    &    1    &  \nodata&    0    &  2  &5\nl
L648-1      &   0   &\nodata& 0     &  \nodata&  \nodata&  \nodata&  \nodata&  0  &2\nl
L673-7      &   0   & 0     & 0     &  \nodata&  \nodata&  \nodata&  \nodata&  0  &3\nl
L694-2      &   1   & 1     & 1     &    1    &    1    &    1    &  \nodata&  6  &6\nl
L1041-2     &   0   & 0     & 0     &  \nodata&  \nodata&  \nodata&  \nodata&  0  &3\nl
L922-1      &   0   & 0     & 0     &  \nodata&  \nodata&  \nodata&  \nodata&  0  &3\nl
L922-2      &   1   & 0     & 0     &  \nodata&  \nodata&  \nodata&  \nodata&  1  &3\nl
L1049-1     &  -1   & 0     & 0     &  \nodata&  \nodata&  \nodata&  \nodata& -1  &3\nl
L1049-2     &   0   & 0     & 0     &  \nodata&  \nodata&  \nodata&  \nodata&  0  &3\nl
L1148       &   0   & 0     & 0     &  \nodata&  \nodata&  \nodata&  \nodata&  0  &3\nl
L1155C-2    &   1   & 0     & 0     &  \nodata&  \nodata&  \nodata&  \nodata&  1  &3\nl
L1155C-1    &   1   & 1     & 0     &    1    &    1    &  \nodata&  \nodata&  4  &5\nl
L1155E      &   0   & 0     & 0     &  \nodata&  \nodata&  \nodata&  \nodata&  0  &3\nl
L1082C-2    &   0   & 0     & 0     &  \nodata&  \nodata&  \nodata&  \nodata&  0  &3\nl
L981-1      &   0   & 0     &-1     &  \nodata&  \nodata&  \nodata&  \nodata& -1  &3\nl
L1063       &   0   & 0     & 0     &  \nodata&    1    &  \nodata&  \nodata&  1  &4\nl
L944-2      &   0   & 0     & 0     &  \nodata&  \nodata&  \nodata&  \nodata&  0  &3\nl
L1251A-2    &   0   & 0     & 0     &  \nodata&  \nodata&  \nodata&  \nodata&  0  &3\nl
L1197       &   0   & 0     & 0     &  \nodata&  \nodata&    1    &  \nodata&  1  &4\nl
L1251       &   0   & 0     & 0     &  \nodata&  \nodata&  \nodata&  \nodata&  0  &3\nl
L1234       &   1   & 1     & 0     &  \nodata&  \nodata&  \nodata&  \nodata&  2  &3\nl
CB246-2     &  -1   &-1     & 0     &   -1    &   -1    &  \nodata&  \nodata& -4  &5\nl
\enddata								          
\tablenotetext{a}{The table gives a summary of line asymmetry colors of starless cores 
with CS (2-1) \& $\rm N_2H^+$ (1-0) observations (LMT99), and CS(3-2) \& $\rm DCO^+$(2-1)
observations (this study).
The column 1 lists the name of source, and 
the columns 2 to 4 illustrate a diagnostics from the normalized velocity differences 
($\delta V$) of CS(2-1), CS(3-2) and DCO$^+$(2-1) with respect to $\rm N_2H^+$ velocities from LMT99.
The numeric numbers `1' or `-1' are given for sources with blue or red
asymmetry of $\vert \delta V\vert \ge 5~\sigma_{\delta V}$, 
and `0' for sources with neither asymmetry of 
$\vert \delta V\vert \le 5~\sigma_{\delta V}$.
The columns 5 to 8 shows other diagnostics for the line asymmetry colors using the 
ratio of intensities of blue component ($\rm T_b$) and red component ($\rm T_r$)
obtained from the Gaussian fit for double-peaked spectra of CS(2-1), CS(3-2), DCO$^+$(2-1),
and $\rm N_2H^+$(1-0) main component.
Here `1', `0', or `-1' are designated for the sources 
of $\rm T_b/T_r > 1+\sigma$, $\rm 1-\sigma \le T_b/T_r \le
1+\sigma$, or $\rm T_b/T_r < 1-\sigma$, respectively, where $\sigma$ is 
one propagation error at each value of the ratio. 
The column 9 gives the algebraic sum of the individual color codes 
of each source, a rating index C. The N in the last column is  
the total number of spectra considered for color codes of each core.}
 
\end{deluxetable}

%\end{document}
% TABLE1.TEX -- AASTeX sample table 1.
 
\special{landscape}
\def\etal{{\it et al.~}}
\setlength{\textwidth}{15cm}
\setlength{\textheight}{20cm}
%\begin{document}
%\addtocounter{page}{}
\begin{deluxetable}{lcccccccc}
%\tablewidth{50pc}
\small
\tablecaption{ Infall Speeds and Optical Depths derived from the two layers fits of CS $(3-2)$ 
and (2-1) lines \tablenotemark{a}}
\tablehead{
        \colhead{} &\colhead{} &\colhead{} &\colhead{} &\colhead{} &
        \colhead{} &\colhead{} \nl
        \colhead{Source} & \colhead{$\tau_{0,21}$ (H)}  &\colhead{$\rm V_{in,21}$(H) }&
        \colhead{$\tau_{0,21}$ (F)}  & \colhead{$\rm V_{in,21}$ (F)}      &
        \colhead{$\tau_{0,32}$ }     & \colhead{$\rm V_{in,32}$(F)} \nl 
        \colhead{} &\colhead{} &\colhead{(km s$^{-1}$)} &\colhead{} &
        \colhead{(km s$^{-1}$)} & \colhead{} &\colhead{(km s$^{-1}$)}
        }
\startdata
L1498         &1.68 & 0.033 &        1.53 &  0.020&  1.21   & 0.023 \nl
L1445         &0.91 & 0.046 &             &       &  0.91   & 0.068 \nl
L1521F        &1.69 & 0.014 &             &       &  1.45   & 0.045 \nl
L1544         &2.44 & 0.020 &        2.71 &  0.015&  1.98   & 0.026 \nl
L183          &1.65 & 0.060 &        1.67 &  0.032&  1.04   & 0.086 \nl
L1689B        &1.46 & 0.037 &        1.73 &  0.037&  1.40   & 0.059 \nl
L158          &1.21 & 0.039 &             &       &  1.52   & 0.066 \nl
L234E-S       &     &       &        1.34 &  0.035&  1.21   & 0.033 \nl
L694-2        &1.69 & 0.051 &        1.92 &  0.077&  1.32   & 0.088 \nl
L1155C-1      &     &       &        1.79 &  0.087&  1.43   & 0.090 \nl
\enddata								          
\tablenotetext{a}{The first column indicates the name of
the source. The second to the 6th columns 
list the optical depths and infall speeds derived from the two layers fit for
the Haystack CS(2-1), the FCRAO CS(2-1), and 12m CS(3-2) lines, respectively. 
The ``21'' and ``32'' designate CS(2-1) and (3-2).
The ``H'' and ``F'' denote Haystack and FCRAO, respectively.
}
 
\end{deluxetable}